\documentclass[a4paper,fleqn,usenatbib]{mnras}

\usepackage{newtxtext,newtxmath}

\usepackage[T1]{fontenc}
\usepackage{ae,aecompl}

\usepackage{graphicx}	
\usepackage{amsmath}	
\usepackage{amssymb}	

\title[Magnetic turbulence around Geminga]{Constraining the properties of the magnetic turbulence in the Geminga region using HAWC $\gamma$-ray data}

\author[L\'opez-Coto, R. \& Giacinti, G.]{
Rub\'en~L\'opez-Coto,$^{1,2}$\thanks{rlopez@pd.infn.it} 
Gwenael~Giacinti $^{1}$\thanks{giacinti@mpi-hd.mpg.de}
\\
$^{1}$Max-Planck-Institut f\"ur Kernphysik, P.O. Box 103980, D 69029 Heidelberg, Germany\\
$^{2}$now at Universit\`{a} di Padova and INFN, I-35131, Padova, Italy
}

\pubyear{2017}

\begin{document}
\label{firstpage}
\pagerange{\pageref{firstpage}--\pageref{lastpage}}
\maketitle

\begin{abstract}
Observations of extended gamma-ray emission around Galactic cosmic-ray (CR) sources can be used as novel probes of interstellar magnetic fields. Using very-high-energy gamma-ray data from the HAWC Observatory, we place constraints on the properties of the magnetic turbulence within $\approx 25$\,pc from Geminga. We inject and propagate individual CR electrons in 3D realizations of turbulent magnetic fields, calculate the resulting gamma-ray emission, and compare with HAWC measurements of this region. We find that HAWC data is compatible with expectations for Kolmogorov or Kraichnan turbulence, and can be well fitted for reasonable coherence lengths and strengths of the turbulence, despite implying a CR diffusion coefficient significantly smaller than those suggested by Galactic CR propagation codes. The best fit is found for a coherence length $L_{\rm c} \approx 1$\,pc and a magnetic field strength $B_{\rm rms} \approx 3\,\mu$G, and the preferred value for $L_{\rm c}$ increases with $B_{\rm rms}$. Moreover, the apparent lack of strong asymmetries in the observed emission allows us to constrain the coherence length to $L_{\rm c} \lesssim 5$\,pc in this region.
\end{abstract}


\section{Introduction}
\label{sec:intro}

Our knowledge of cosmic-ray (CR) origin and propagation in the Galaxy mainly comes from CR measurements performed by balloons, satellites or air shower arrays at Earth, from gamma-ray observations (both hadronic and leptonic in origin), and from the Galactic synchrotron emission. For a review on CR propagation and interactions, see~\cite{Strong07}. In particular, gamma-ray observations have been very successful at improving our knowledge of CR origin ---e.g.~\cite{Ackermann13,hess16}, and propagation in the disk~\citep{Ackermann11} and halo~\citep{Tibaldo15} of the Milky Way. CR propagation around their sources can be studied too in gamma rays, by looking for the emission (or lack thereof) from nearby molecular clouds, see e.g.~\cite{Gabici2010,Gabici2013,Cui2016}. Such studies often assume isotropic diffusion of CRs around their sources and deduce the corresponding CR diffusion coefficient.

However, as suggested in~\cite{2012PhRvL.108z1101G,2013PhRvD..88b3010G}, more information could be retrieved from observations of extended gamma-ray emission around CR sources, such as the coherence length of the turbulence surrounding the source. These localized constrains on interstellar magnetic fields could bring, in the future, complementary information to that gained from radio observations. The coherence length of the interstellar turbulence typically lies in the range $L_{\rm c} \sim 1 - 100$\, pc, depending on location~\citep{Haverkorn:2008tb,Iacobelli2013}. This is significantly larger than the gyroradii of Galactic CRs. CRs escaping from their sources then follow magnetic field lines in the surrounding medium, resulting in an extended gamma-ray emission that must be asymmetric as long as the bulk of escaping CRs is at distances smaller than a few $L_{\rm c}$ from the source~\citep{2012PhRvL.108z1101G,Malkov2013,Kistler2012,Nava2013}. This could be the reason for the bipolar structure of the GeV gamma-ray emission observed by Fermi-LAT~\citep{Uchiyama2012} around the old supernova remnant W44~\citep{Malkov2013}. On the other hand, when the bulk of emitting CRs is beyond a few $L_{\rm c}$ from the source, field lines appear tangled on these scales and the emission is expected to look more symmetric, provided the turbulence level is high. Detection of a symmetric extended emission would constrain the local value of $L_{\rm c}$ to be smaller than a fraction of the size of the emitting region.

An interesting example of extended gamma-ray emission with a leptonic origin is that surrounding the Geminga pulsar. Geminga (PSR J0633+1746) was the first pulsar discovered in gamma rays and the second brightest steady source at GeV energies observed to date. It has a spin-down power of $\dot{E}=3.26\times10^{34}$\,erg/s, a period $P=0.237$\,s, a period derivative $\dot{P}=1.1\times10^{-14}$\,s~s$^{-1}$ and a characteristic age $\tau\sim3.4\times10^5$\,yr~\citep{Bertsch92}. Geminga is one of the nearest pulsars, located at a distance $d=250^{+230}_{-80}$\,pc from Earth~\citep{Verbiest12,Faherty07}. Very-high-energy (VHE) gamma rays coming from the region surrounding Geminga were discovered by the Milagro experiment \citep{Milagro_Geminga}. The angular size reported for the source was $\sim2^\circ$, challenging to be detected for subsequent imaging atmospheric Cherenkov telescopes~\citep{Ahnen16}. HAWC reported in \cite{Geminga_hawc} the detection of VHE $\gamma$-ray emission surrounding Geminga and another nearby pulsar PSR B0656+14. Using the VHE surface brightness measured for both sources, they determined that electrons and positrons (hereafter electrons) were transported via diffusion outside of these sources and measured a diffusion coefficient of $D_{100} = (4.5 \pm 1.2 ) \times 10^{27}$\,cm$^2$\,s$^{-1}$ for $\sim 100$\,TeV electrons. This diffusion coefficient is more than two orders of magnitude lower than that derived from measurements of the Boron-to-Carbon ratio within standard assumptions~\citep{Strong07,Aguilar16}. 

In the present paper, we investigate which constraints can be placed with the data of~\cite{Geminga_hawc} on the properties of the turbulence within a $\approx 25$\,pc radius region around the Geminga pulsar, the diffusion radius measured in ~\cite{Geminga_hawc}. We only use the data from Geminga, because it has the highest flux amongst the two extended sources detected by HAWC. We calculate from first principles the individual trajectories of very-high-energy electrons in three-dimensional realizations of magnetic turbulence. This allows us to study effects that cannot be accounted for with the standard electron diffusion-loss equation used in~\cite{Geminga_hawc}, such as non-diffusive and, or, highly anisotropic propagation on scales $\lesssim L_{\rm c}$. Also, our approach automatically provides self-consistent calculations of the electron diffusion coefficient and energy losses, while relating them to the exact power-spectrum of the turbulence we choose to study. 

In Section~\ref{Simulations}, we present the numerical simulations we perform. Our results and their comparison with HAWC data are shown in Section~\ref{Results}. We discuss these results in Section~\ref{Discussion} and conclude in Section~\ref{Conclusions}.

\section{Numerical Simulations}
\label{Simulations}

In this Section, we describe how we produce the synthetic gamma-ray surface brightness maps that are compared with HAWC data in Section~\ref{Results}.

\subsection{Propagation of VHE electrons}

Instead of describing electron propagation with the standard diffusion-loss equation, we calculate from first principles the individual trajectories of VHE electrons in 3D realizations of magnetic turbulence: we propagate the electrons, using the Lorentz force and taking into account their energy losses due to synchrotron and inverse Compton. We use such a procedure for two main reasons. First, this allows us to relate directly the gamma-ray surface brightness to the actual parameters of the magnetic turbulence that is probed by the electrons: Its power-spectrum $\mathcal{P}(k)$, coherence length $L_{\rm c}$ (or outer scale $L_{\max}$), and the root-mean-square strength of the magnetic field $B_{\rm rms} \equiv \sqrt{\langle B^{2} \rangle}$. In particular, this allows us to calculate electron propagation and energy losses self-consistently for any given set of turbulence parameters. Second, this enables us to decribe accurately effects which cannot be studied within the diffusion approximation, and which can play a role close to the source: Non-diffusive propagation at very early times, and highly anisotropic propagation on $\sim L_{\rm c}$ length-scales due to electrons gyrating along, and following, magnetic field lines. The latter effect is expected to produce filamentary or irregular structures in the $\gamma$-ray emission close to the source, when the bulk of particles is still contained in a rather well-defined magnetic flux tube encompassing the source, and highlights local field lines. See \cite{2012PhRvL.108z1101G} and~\cite{2013PhRvD..88b3010G} for more details. As will be seen in Section~\ref{Results}, taking these effects into account enables us to put relevant constraints on $L_{\rm c}$.

The measured $\gamma$-ray spectrum of Geminga follows a power-law $dN/dE = f_0 (E/E_0)^{-\Gamma}$ between 8 and 40\,TeV, with $f_0= 1.36 \times 10^{-14}$\,TeV$^{-1}$\,cm$^{-2}$\,s$^{-1}$, $E_0=20$\,TeV and $\Gamma=2.34$, see \cite{Geminga_hawc}. The bulk of the emission observed by HAWC is therefore due to $\sim 100$\,TeV electrons. Because of their short cooling time at such energies ($\sim$~a few tens of kyr), we make the reasonable assumption that Geminga has injected them steadily on such time scales, with a time-independent power-spectrum. Taking into account electron cooling and integration of the photon emission over the line of sight, the observed $\gamma$-ray spectral index is consistent with the electrons being injected with a power-law spectrum $dN/dE = f_{\rm e} (E/E_0)^{-\alpha}$ with $\alpha=2.24$~\citep{Geminga_hawc}.

We propagate 5000 electrons in every tested realization of magnetic turbulence (620 realizations in total ---details below). Considering Geminga to be point-like, we inject these electrons at a given point in space, in the simulation. We distribute their initial energies between 40\,TeV to 500\,TeV according to a power-law spectrum $dN/dE$ with index 2.24. Injecting them continuously in time and waiting for the steady state regime to appear would be inefficient and prohibitive in terms of computing time. Instead, we inject all electrons at $t=0$, and record their coordinates (energy and position in space) at time intervals equally spaced by $\Delta t$. One can then consider each recording as a new emitting particle for the calculation of the $\gamma$-ray emission. We verified that $\Delta t = 20$\,yr produces correct results. Since we are only interested in $\gamma$-rays with energies above 8\,TeV, we stop propagating electrons once their energies drop below 39\,TeV. Electron trajectories are calculated within the continuous energy-loss approximation. We take into account synchrotron and inverse Compton losses, which are dominant at these energies as shown in \cite{Diffusion_electrons_paper}. Klein-Nishina effects on their inverse Compton scattering of CMB photons are not negligible at the highest energies we consider here. We use the approximation presented in~\cite{Moderski05}, and the energy loss per time unit $|dE/dt|$ of an electron with energy $E$ reads:
\begin{align}
\left| \frac{dE}{dt} \right| \simeq & \, 2.53 \times 10^{-15} \, {\rm TeV/s} \, \bigg[ \left( \frac{B}{\mu{\rm G}} \right)^{2} + 10.1
\nonumber\\
& \times \left( 1 + \frac{E}{99\,{\rm TeV}} \right)^{-1.5} \, \bigg] \left(\frac{E}{{\rm TeV}}\right)^{2}\;,
\end{align}
where $B$ is the magnetic field strength at the considered point on the particle trajectory.

The turbulent magnetic field is generated on a set of nested grids, using the method presented and tested in \cite{Giacinti2012}. See Section~2 of~\cite{Giacinti2012} for a detailed description, and Section~2 of~\cite{2013PhRvD..88b3010G} for a justification of the underlying simplifications, such as the use of a static field. In the following, we use (isotropic) Kolmogorov turbulence ($\mathcal{P}(k) \propto k^{-5/3}$) and Kraichnan turbulence ($\mathcal{P}(k) \propto k^{-3/2}$), on the interval $2\pi/L_{\max} \leq k \leq 2\pi/L_{\min}$. $L_{\min}$ is set to about a tenth of the electron gyroradius at the lowest energies we consider (39\,TeV), so that all particles experience resonant scattering. We verified that taking smaller values for $L_{\min}$ would not affect our results. The outer scale of the turbulence, $L_{\max}$, is a free parameter which we vary. In the following, we present our results in terms of the coherence length, which is given by $L_{\rm c} \simeq L_{\max}/5$ for Kolmogorov turbulence and $L_{\rm c} \simeq L_{\max}/6$ for Kraichnan turbulence, cf. Eq.~(2.1) in \cite{Giacinti2012}. In principle, the power-spectrum of the interstellar turbulence may be different from these two examples, and may be anisotropic. However, the current data from HAWC does not require any such refinement to our treatment yet, because isotropic turbulence fits it well for reasonable values of $L_{\rm c}$ and $B_{\rm rms}$, see Section~\ref{Results} and discussion in Section~\ref{Discussion}.

For both power-spectra, we do calculations for 9 different values of $L_{\rm c}$: 0.1, 0.25, 0.5, 1, 2.5, 5, 10, 20, and 40\,pc. For each of these cases, we also test 3 or 4 different turbulent magnetic field strengths: $B_{\rm rms} = 2$, 3, 4, and $5\,\mu$G for $L_{\rm c} \leq 1$\,pc, and $B_{\rm rms} = 3$, 4, and $5\,\mu$G for $L_{\rm c} \geq 2.5$\,pc. For every studied combination of turbulence parameters $\{\mathcal{P}(k),L_{\rm c},B_{\rm rms}\}$, we randomly generate 10 different realizations of turbulence with those same parameters. While each realizations must be separately compared and fitted to HAWC data, we also provide, for each combination of parameters $\{\mathcal{P}(k),L_{\rm c},B_{\rm rms}\}$, the median results (e.g. $\chi^{2}/n.d.f.$) and confidence intervals over the ten corresponding realizations. This allows us to take into account the fluctuations due to \lq\lq cosmic variance\rq\rq\/ from one realization to another. As expected, these fluctuations are negligible for values of $L_{\rm c}$ that are very small with respect to the size of the $\gamma$-ray emitting region, but become important for larger values of $L_{\rm c}$.

\subsection{Resulting gamma-ray emission}


To compute the VHE $\gamma$ ray spectrum produced by the electrons after diffusing away from the central source, we used libraries from \texttt{edge} and \texttt{gamera} \citep{Diffusion_electrons_paper, GAMERA}. The $\gamma$-ray spectrum produced at the energies considered is dominated by the upscattering of electrons into CMB photons, being the only ones considered as photon targets in the VHE $\gamma$-ray calculation. The electron-photon interaction is calculated using the full Klein-Nishina treatment of the cross section \citep{Blumenthal70}. We also use \cite{Blumenthal70} to calculate the emissivity of the radiation fields and integrate it over the photon and electron spectra. The radiation target is added as a grey body distribution with mean the temperature of the CMB.
Since the number of simulated electrons is smaller than the one needed to produce HAWC's spectrum, to compare our calculation to HAWC's result, we normalize the total VHE $\gamma$-ray emission produced in our model to the one measured by HAWC. We then compute the normalized surface brightness obtained for different propagation assumptions as a function of the angular distance from the source.

\section{Results}
\label{Results}

\subsection{Diffusion coefficient at 100\,TeV}
\label{D_at_100TeV}

\begin{figure*}
\begin{center}
\includegraphics[width=0.49\textwidth]{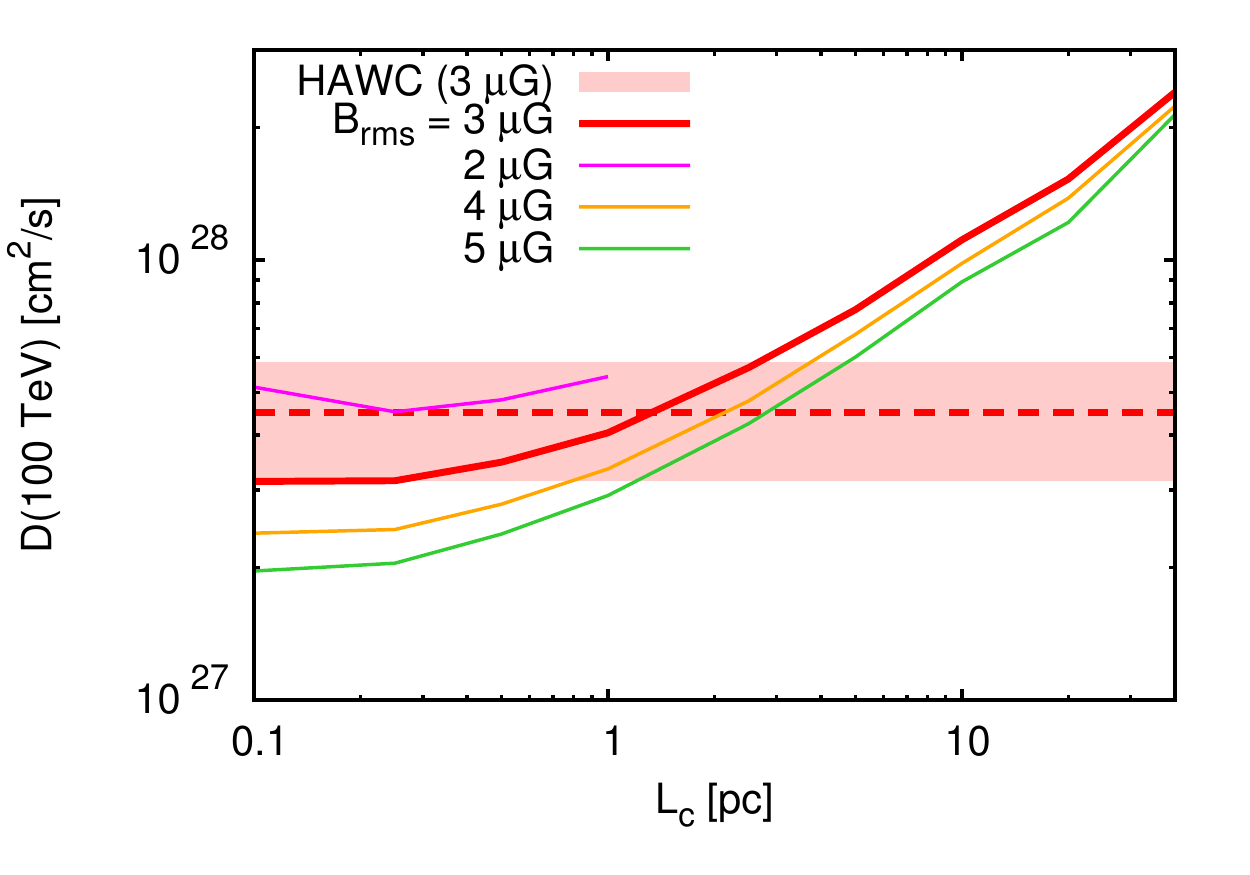}
\includegraphics[width=0.49\textwidth]{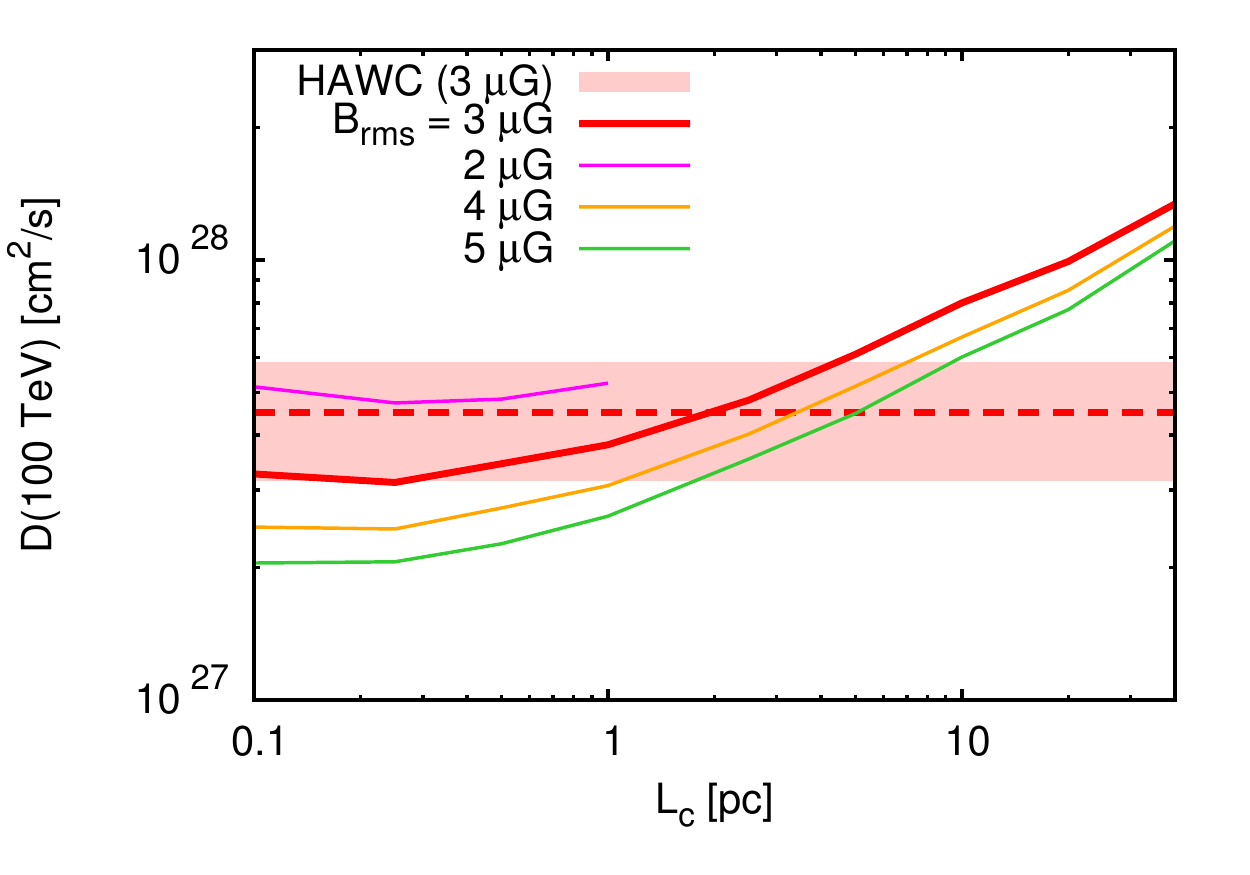}
\caption{Diffusion coefficient of 100\,TeV CRs in pure Kolmogorov (left panel) and Kraichnan turbulence (right panel), versus the turbulence coherence length $L_{\rm c}$. Each solid line corresponds to a different magnetic field strength: $B_{\rm rms}=3\,\mu$G (thick red line), 2\,$\mu$G (magenta), 4\,$\mu$G (orange), and 5\,$\mu$G (green). HAWC measurement assuming $B_{\rm rms}=3\,\mu$G~\citep{Geminga_hawc} is shown with the dotted red line, together with its uncertainty (pale red area).}
\label{fig:diff_coef}
\end{center}
\end{figure*}

Before discussing our main results for the $\gamma$-ray surface brightness profiles, it is instructive to calculate the CR diffusion coefficient at 100\,TeV. HAWC Collaboration measured it as $D_{100} = (4.5 \pm 1.2 ) \times 10^{27}$\,cm$^2$\,s$^{-1}$~\citep{Geminga_hawc}. Since the emission around Geminga is approximately symmetric, there is no strong regular magnetic field in that region, unless it happens to be aligned with the line of sight. The latter possibility is not likely, because regular fields in the Galactic disk are thought to follow spiral arms, and the direction to Geminga is not aligned with that of the Orion Spur. Therefore, we calculate the isotropic diffusion coefficient in pure turbulence. To do so, we propagate $10^{4}$ protons with 100\,TeV energy, calculate their instantaneous diffusion coefficient $D(t)$, and take its limit $D$ at large $t$. For each set of physical parameters of the turbulence, we use 10 magnetic field realizations, and thus $10^{3}$ protons per configuration. The results for Kolmogorov (resp. Kraichnan) turbulence are presented in the left (resp. right) panel of Figure~\ref{fig:diff_coef}, for $L_{\rm c}$ in the range $0.1-40$\,pc. Note that 0.1\,pc is somewhat smaller than the values of coherence lengths usually thought to be relevant for interstellar magnetic fields. HAWC measurement for $B_{\rm rms}=3\,\mu$G is shown with the red dotted line, and its uncertainty with the pale red area. Each solid line corresponds to a different strength of the turbulence: Thick red line for $B_{\rm rms}=3\,\mu$G (to be compared with HAWC data), magenta for 2\,$\mu$G, orange for 4\,$\mu$G, and green for 5\,$\mu$G. As expected, the diffusion coefficient $D$ is smaller for larger values of $B_{\rm rms}$. At $L_{\rm c} \gtrsim 1$\,pc, one probes the regime where the Larmor radius $r_{\rm g}$ of 100\,TeV CRs in a few $\mu$G field is $\ll L_{\rm c}$, and where $D \propto L_{\rm c} (r_{\rm g}/L_{\rm c})^{2-\delta}$ for $\mathcal{P}(k) \propto k^{-\delta}$. See e.g. \cite{Casse2002,Aloisio2004}. The expected limiting slope for $D(L_{\rm c})$ starts to be reached at the largest values of $L_{\rm c}$ considered here. The change of slope at $L_{\rm c} \lesssim 1$\,pc is due to the fact that $L_{\rm c}$ is not very small compared to $r_{\rm g}$ there: In this region of parameter space, the diffusion coefficient slowly begins its transition towards the $D \propto L_{\rm c} (r_{\rm g}/L_{\rm c})^{2}$ behaviour reached at $r_{\rm g} \gg L_{\rm c}$.

By comparing the thick red lines in Figure~\ref{fig:diff_coef} with HAWC measurements, one can see that values of $L_{\rm c}$ lying in range $\approx 0.3 - 3$\,pc (resp. $\approx 0.3 - 5$\,pc) are compatible with the data, within the quoted error-bars, for Kolmogorov (resp. Kraichnan) turbulence with $B_{\rm rms}=3\,\mu$G. Also, the best fit is reached for $L_{\rm c} \approx 1$\,pc (resp. $L_{\rm c} \approx 2$\,pc) for Kolmogorov (resp. Kraichnan). These values of coherence lengths are realistic, and are close to the lower end of the range of relevant values for the interstellar medium. The lines for 2, 4, and 5\,$\mu$G cannot be directly compared with the data, because the value of the diffusion coefficient is not published for these values of $B_{\rm rms}$. These cases will be studied in the next Subsection by comparing our predicted surface brightness profiles with that measured by HAWC Collaboration.

\begin{figure*}
\begin{center}
\includegraphics[width=0.32\textwidth]{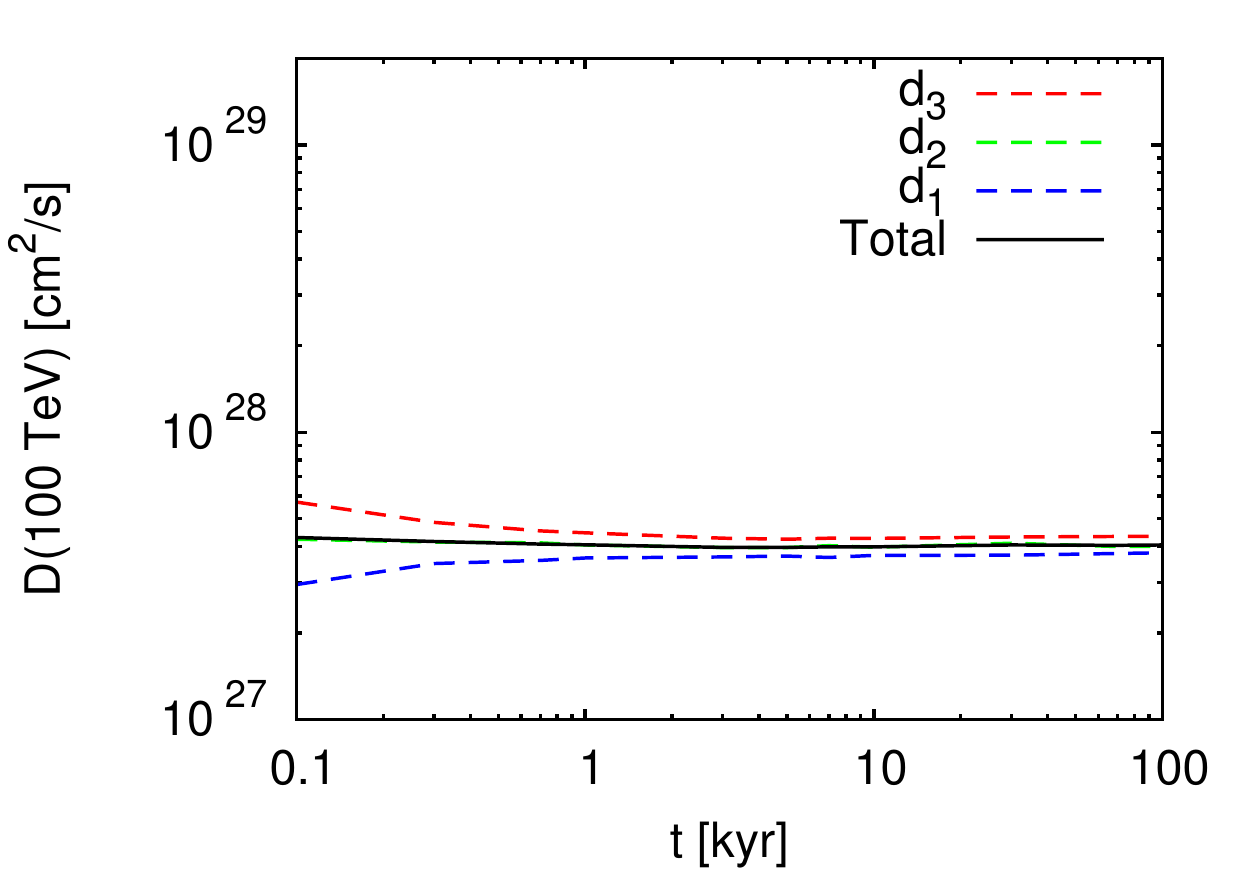}
\includegraphics[width=0.32\textwidth]{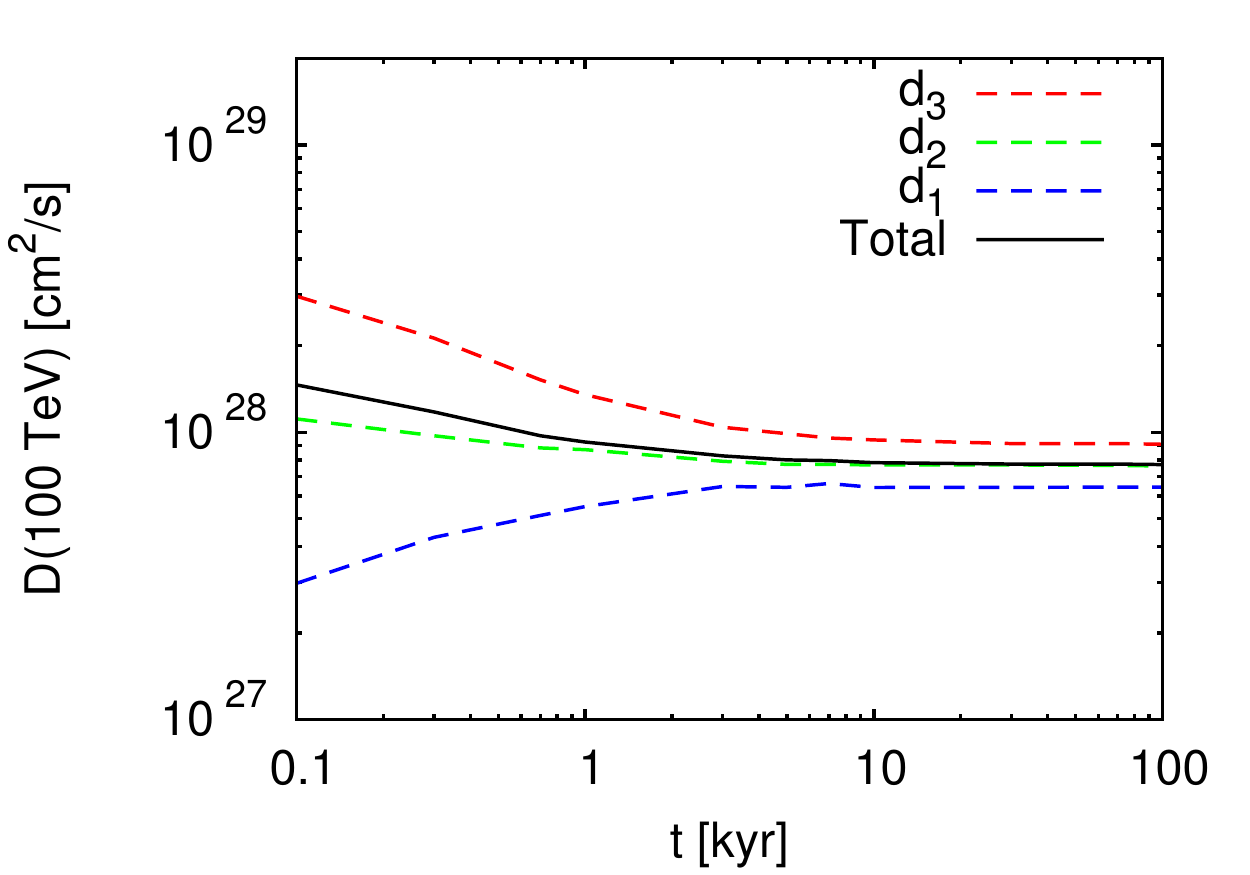}
\includegraphics[width=0.32\textwidth]{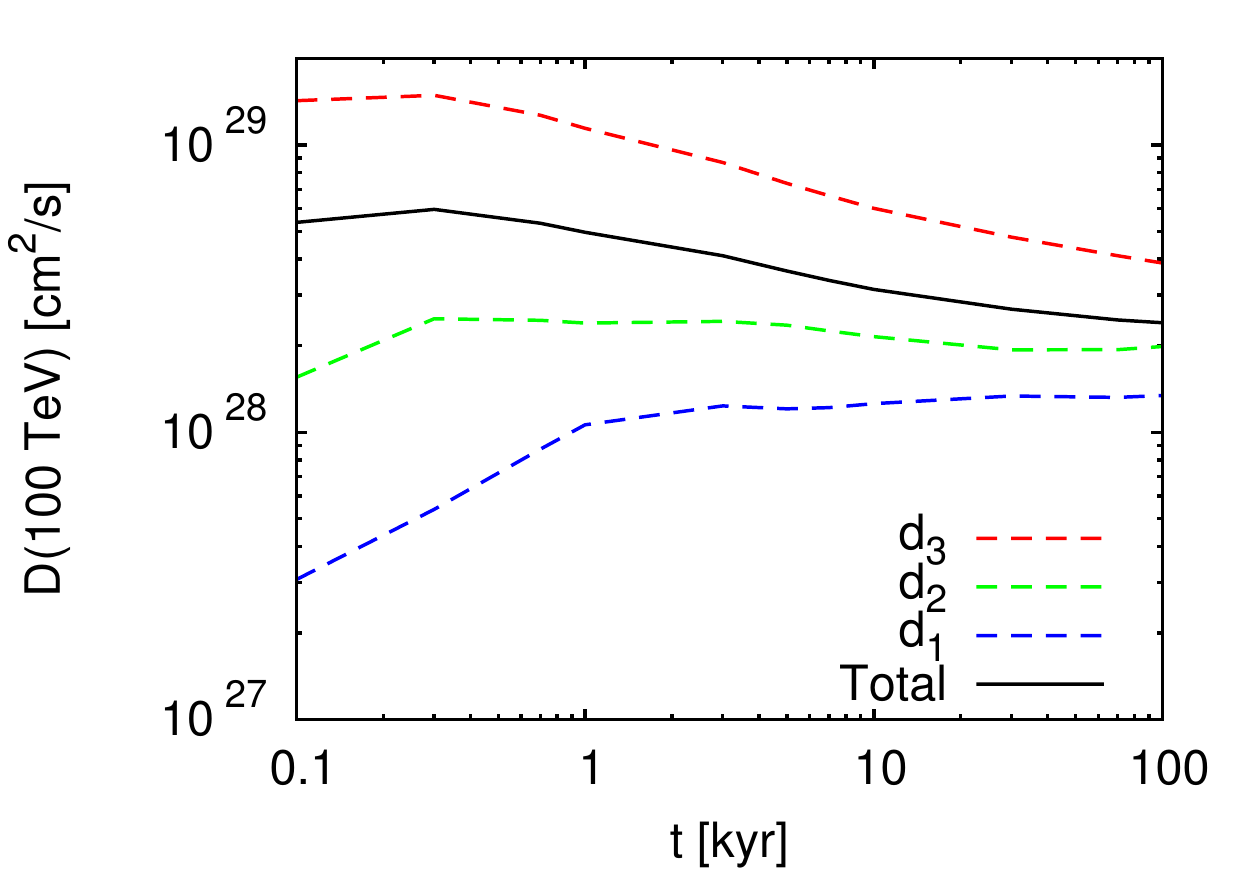}
\caption{Eigenvalues $d_{1,2,3}$ (blue, green and red dashed lines) of the diffusion tensor $D_{ij}=\langle x_ix_j\rangle/(2t)$ for 100\,TeV protons, as a function of their propagation time $t$ in Kolmogorov turbulence with $B_{\rm rms}=3\,\mu$G. $d_{1,2,3}$ are averaged over 10 different realizations of turbulence. $L_{\rm c} = 1$\,pc (left panel), 5\,pc (centre), and 40\,pc (right). Black solid line for the average instantaneous diffusion coefficient as a function of $t$.}
\label{fig:eigenvalues}
\end{center}
\end{figure*}

Another important property is the asymmetry (or lack thereof) of the distribution of CRs around their sources. To estimate this, we inject 100\,TeV protons at $(x,y,z)=(0,0,0)$ and $t=0$, and calculate the eigenvalues $d_{1,2,3}^{(k)}$ of the diffusion tensor $D_{ij}(t)=\langle x_ix_j\rangle/(2t)$ in each magnetic field realization $(k)$, where $k=1,...,10$. In Figure~\ref{fig:eigenvalues}, we present the average eigenvalues $d_{1,2,3}=\sum_{k=1}^{10}d_{1,2,3}^{(k)}/10$ as a function of $t$, with blue, green and red dashed lines, see key for the ordering $d_{1}<d_{2}<d_{3}$. We use Kolmogorov turbulence with $L_{\rm c}=1$\,pc, 5\,pc and 40\,pc, respectively for the left, centre and right panels. $B_{\rm rms}=3\,\mu$G in all three panels. The solid black lines represent the average instantaneous diffusion coefficients $D(t)$. At sufficiently long times $t$ after CR injection, the diffusion regime is reached and $D(t)$ plateaus at its limiting value $D$. As can be seen by comparing the three panels in Fig.~\ref{fig:eigenvalues}, reaching this plateau takes more time for larger coherence lengths. The spread between the three eigenvalues is initially large, and decreases with time up until $d_{1}$, $d_{2}$, and $d_{3}$ all converge towards the same limit $D$. This convergence takes more time for larger values of $L_{\rm c}$, too. The spread between the eigenvalues at early times shows that particle propagation proceeds significantly faster in some directions than in others. This difference can be orders of magnitude: For example, there are two orders of magnitude difference between $d_{1}$ and $d_{3}$ at $t=0.1$\,kyr for $L_{\rm c}=40$\,pc. This is due to the fact that particles follow the magnetic field lines surrounding their point of injection, i.e. their source. Their propagation is then initially very anisotropic. As long as the bulk of particles remains at distances $\lesssim$~a few $L_{\rm c}$ from the source (early times $t$), the resulting $\gamma$-ray emission for an observer at Earth appears asymmetric. The convergence of the eigenvalues at late times shows that CR propagation tends towards isotropic diffusion. This happens when CRs reach distances $\gtrsim$~a few $L_{\rm c}$ form the source. This is why the eigenvalues take more time to converge for larger $L_{\rm c}$. Propagation of each individual CR still remains anistropic at late times, because they all follow their own local field lines, but the tangling of field lines on scales $\approx L_{\rm c}$ makes the CR distribution around the source appear symmetric, as a whole.

The cooling time of $\sim 100$\,TeV electrons is $\sim 10$\,kyr. By looking at the separation between the eigenvalues at $t \approx 10$\,kyr in Fig.~\ref{fig:eigenvalues}, one can infer that the $\gamma$-ray emission around Geminga should look symmetric for $L_{\rm c} = 1$\,pc, but very asymmetric for $L_{\rm c} = 40$\,pc.

\subsection{Gamma-ray emission: Surface brightness profiles, and best fits to the data}

\begin{figure*}
\begin{center}
\includegraphics[width=0.48\textwidth]{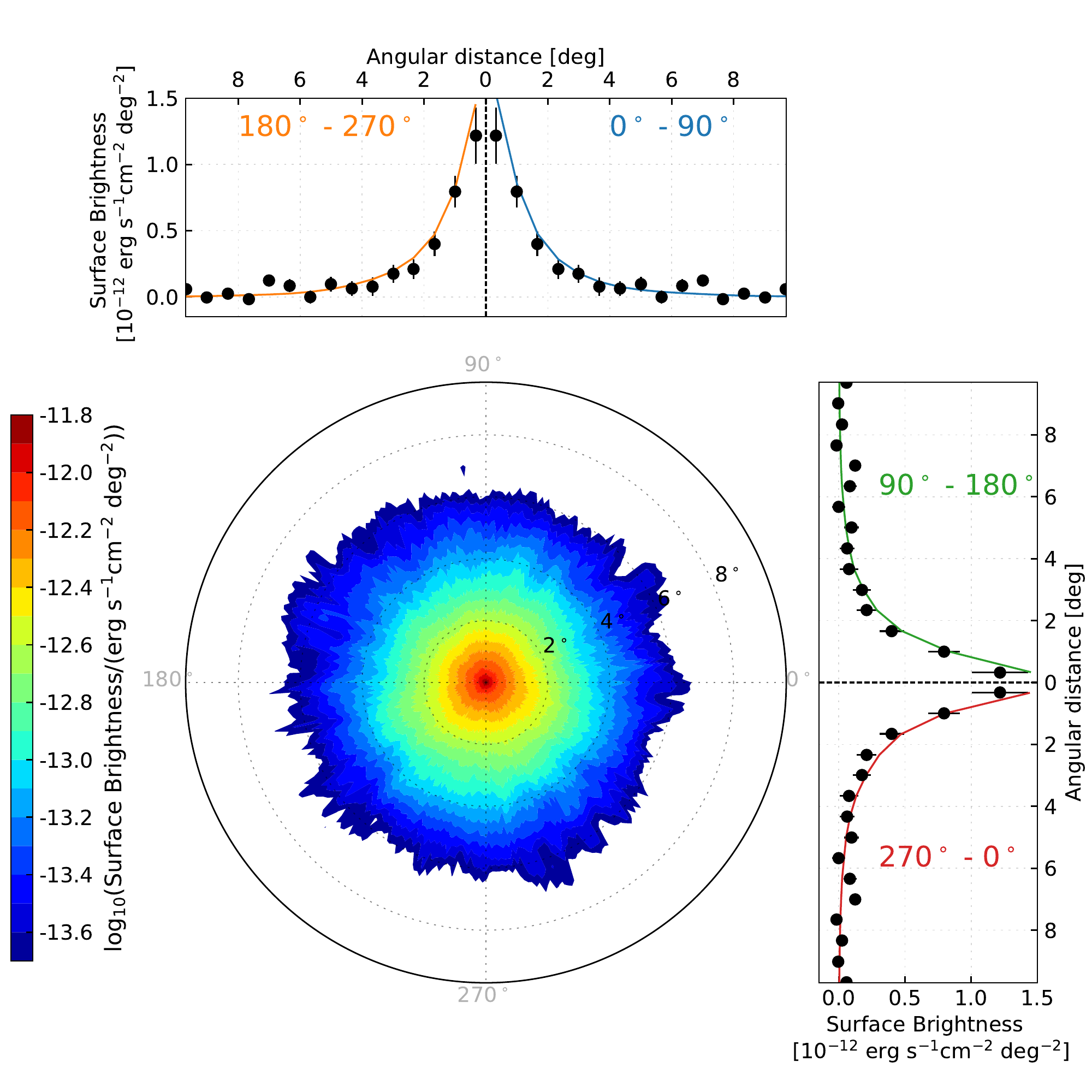}
\includegraphics[width=0.48\textwidth]{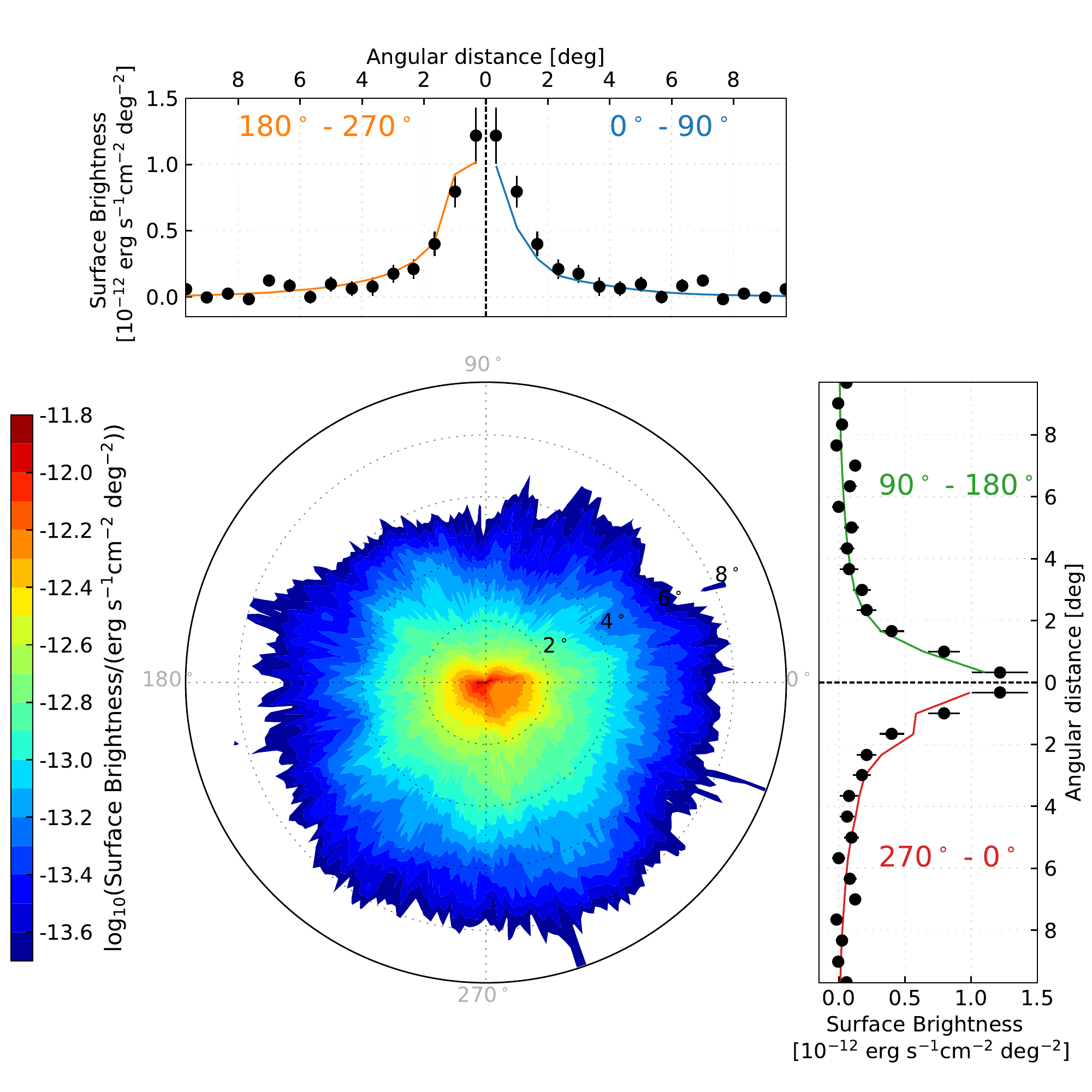}
\includegraphics[width=0.48\textwidth]{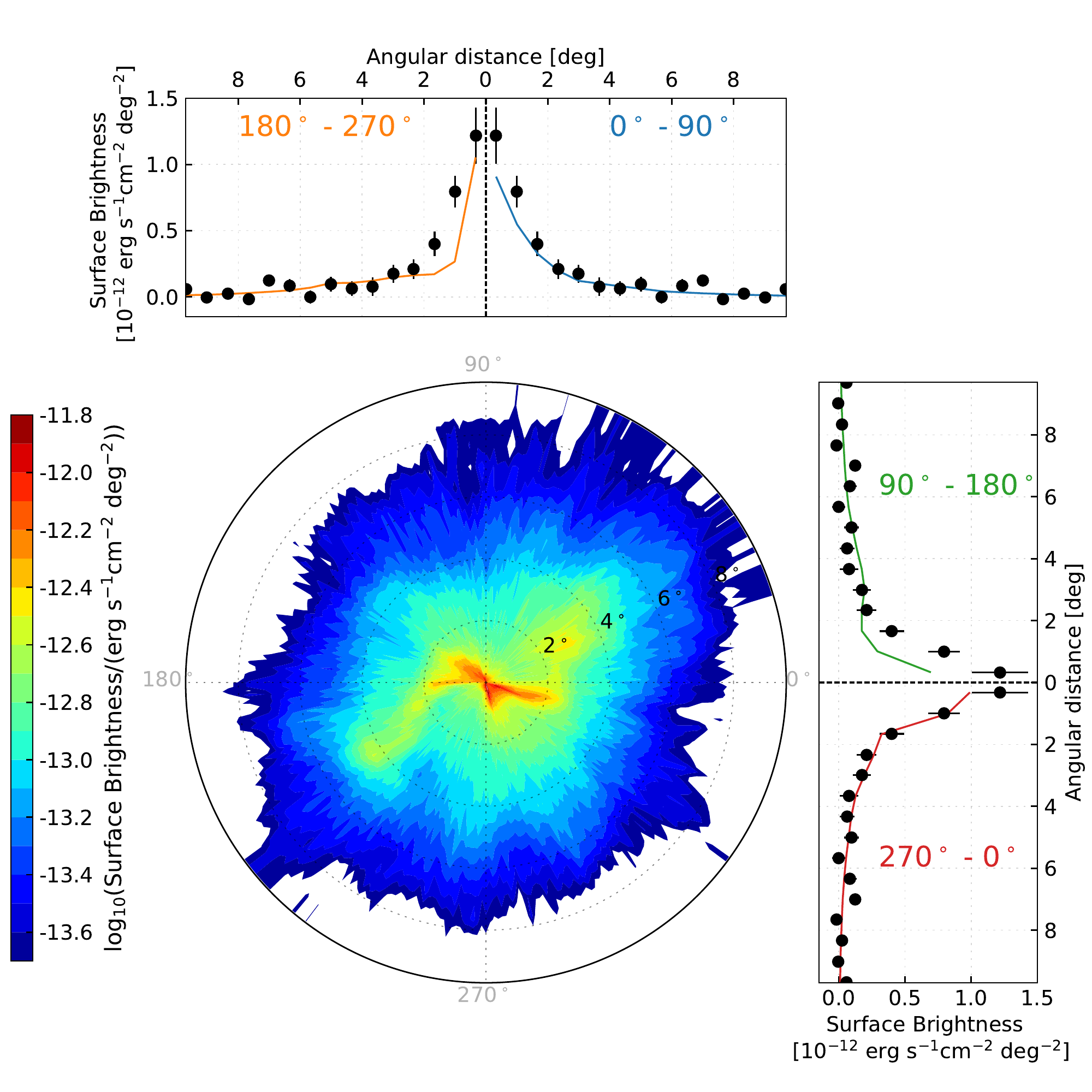}
\includegraphics[width=0.48\textwidth]{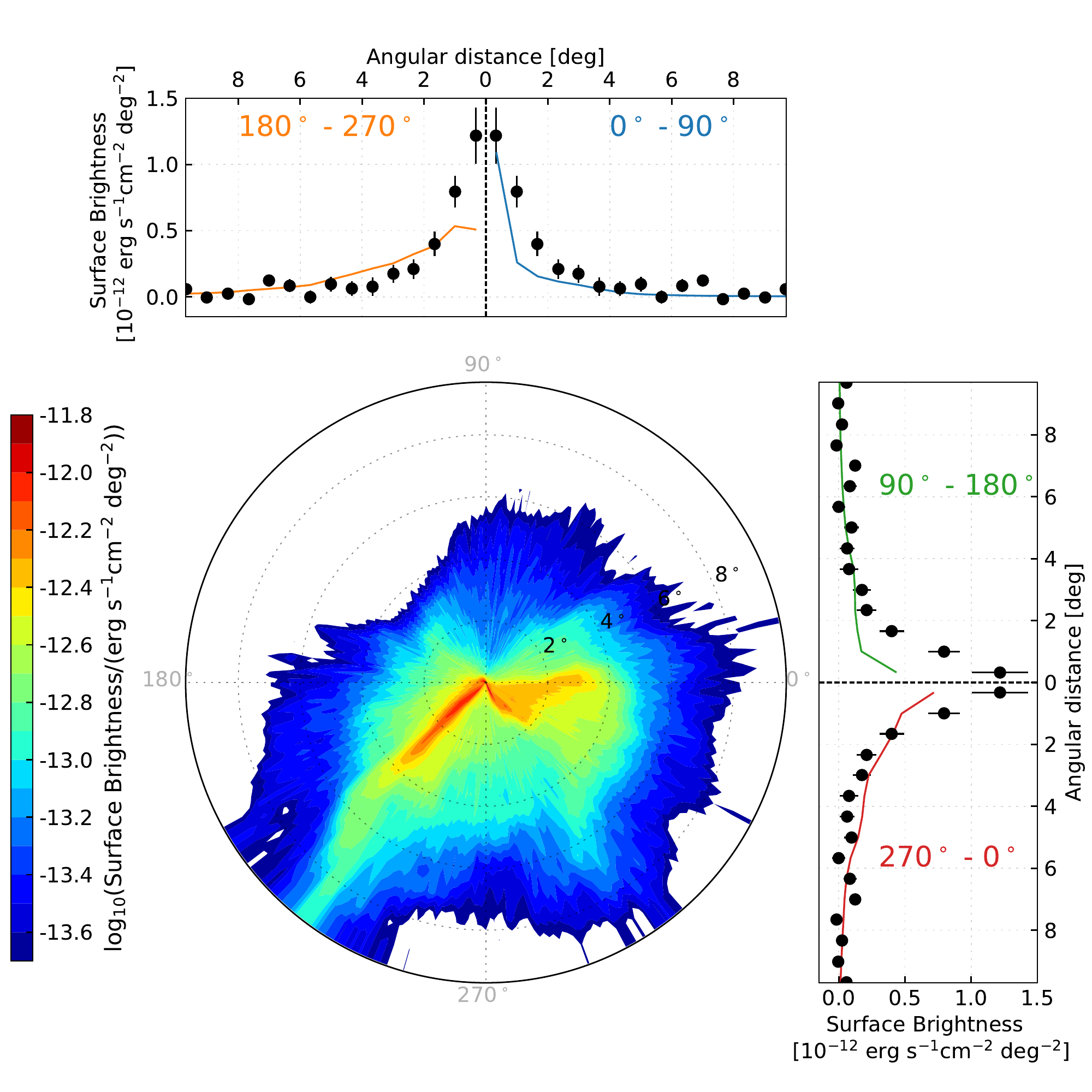}
\caption{VHE $\gamma$-ray surface brightness in polar coordinates for different coherence lengths. $L_{\rm c} = 0.25$\,pc (top left), $L_{\rm c} = 5$\,pc (top right), $L_{\rm c} = 10$\,pc (bottom left), $L_{\rm c} = 40$\,pc (bottom right). The radial distance indicates the angular distance from the pulsar and is marked with black ticks. The polar angle is marked with grey ticks. The surface brightness integrated over a quadrant is located in every panel at the top and right side of every plot in polar coordinates. The color line in these plots corresponds to the result of our calculation in that quadrant, while the black points are the HAWC measured VHE gamma-ray surface brightness.}
\label{fig:2Dprofiles}
\end{center}
\end{figure*}

Let us compare the $\gamma$-ray surface brightness obtained from our simulations with those reported by HAWC for Geminga. Figure~\ref{fig:2Dprofiles} illustrates the evolution of the asymmetry in the surface brightness for different $L_{\rm c}$. We show in polar coordinates the surface brightness as seen in the sky. The plots are shown for Kolmogorov turbulence, a magnetic field with $B_{\rm rms}=3\,\mu$G and coherence lengths $L_{\rm c}=0.25, 5, 10$ and $40$\,pc (see Figure caption for details). The maximum of each panel is marked by the colorscale shown at the left side of it. We can see that for $L_{\rm c}>5$\,pc, the VHE $\gamma$-ray surface brightness distribution is very anisotropic, due to the electrons propagating along the magnetic field lines. At the top and the right side of each figure, we show the surface brightness integrated over the corresponding quadrant (stated in the inset of the plot), and compared to HAWC measured radial profile for Geminga. The surface brightness in the four quadrants become more similar for small coherence lengths due to the isotropy of the distribution of electrons in these examples. We can see that for $L_{\rm c}=0.25$\,pc, the emission is isotropic and very peaked at the center, with an overshoot of the VHE $\gamma$-ray emission close to the center, while for $L_{\rm c}=40$\,pc the opposite occurs, with a more irregular profile due to the preferred direction where electrons propagate in a turbulent magnetic field with such a coherence length.
Before determining which coherence length provides the highest goodness of fit to HAWC's measured surface brightness, we can qualitatively state that for coherence lengths $L_{\rm c}>5$\,pc, the surface brightness produced by our model does not give a good fit to HAWC measurements. The reason is connected to the non-isotropic diffusion of electrons on the length scale of the emission, which results in an irregular distribution of its surface brightness, cf. the plots in polar coordinates.

\begin{figure*}
\begin{center}
\includegraphics[width=0.8\textwidth]{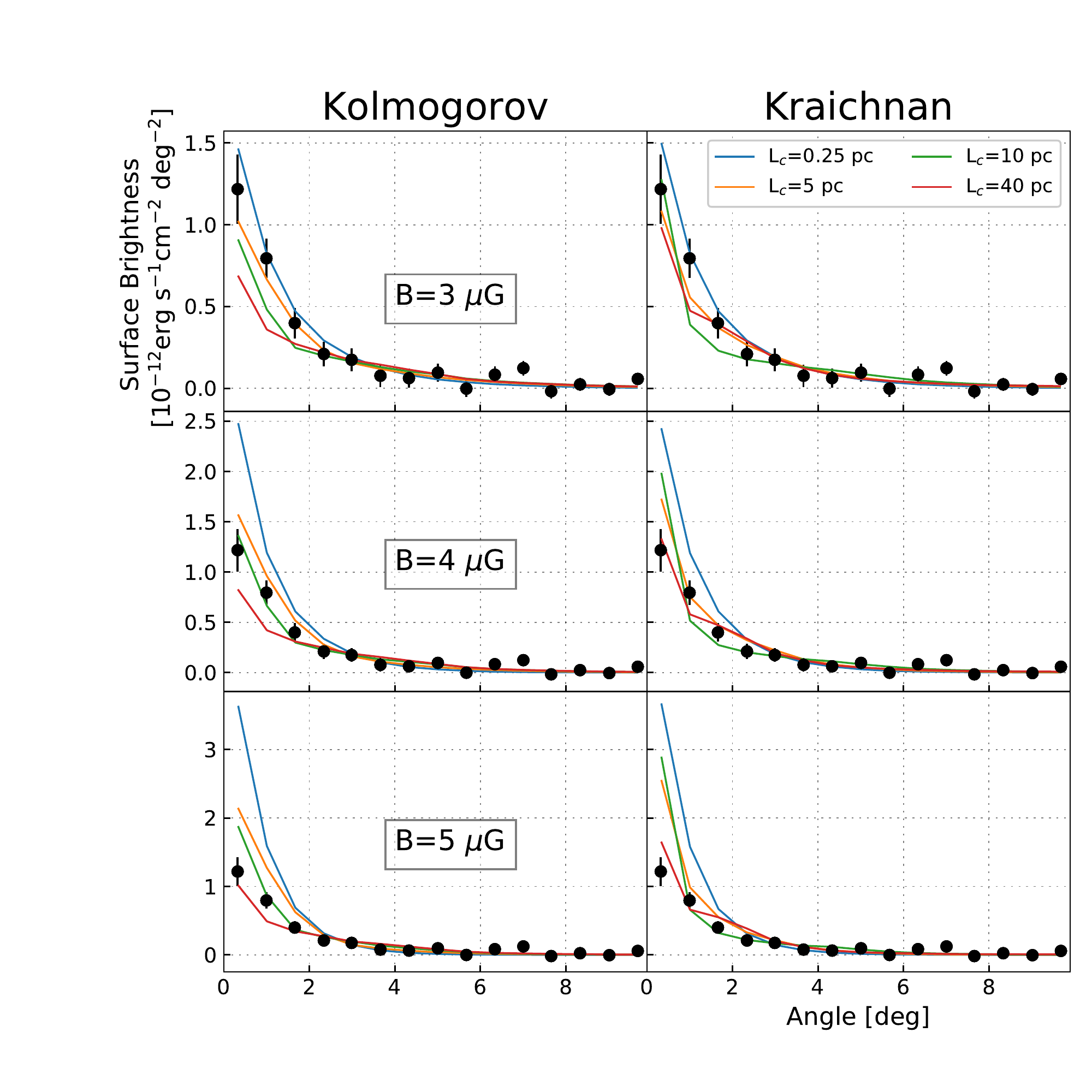}
\caption{VHE $\gamma$-ray surface brightness as a function of the angular distance from the pulsar, for Kolmogorov (left panels) and Kraichnan (right panels) turbulence. The black points represent the HAWC VHE gamma-ray surface brightness. The top panels correspond to a magnetic field strength $B_{\rm rms} = 3\,\mu$G, the middle panels to $B_{\rm rms} = 4\,\mu$G, and the bottom panels to $B_{\rm rms} = 5\,\mu$G. Each line colour corresponds to a different coherence length: Blue for $L_{\rm c} = 0.25$\,pc, orange for $L_{\rm c} = 5$\,pc, green for $L_{\rm c} = 10$\,pc and red for $L_{\rm c} = 40$\,pc.}
\label{fig:1Dprofiles}
\end{center}
\end{figure*}

To perform a quantitative comparison between the surface brightness of our model and HAWC data, we compute the surface brightness integrated over all the azimuthal angles. In Figure \ref{fig:1Dprofiles}, for each set of turbulence parameters (power spectrum, magnetic field strength, and coherence length), we show the results for one given realization of the turbulence. We can confirm that for the same magnetic field strength, the VHE $\gamma$-ray surface brightness profiles are more peaked for smaller coherence lengths. The reason is that the diffusion coefficient increases with $L_{\rm c}$ and therefore particles propagate further before cooling down. It is also visible that for increasing magnetic field strength, the best fit results are moving towards higher coherence length. There is a slight difference between the considered Kolmogorov and Kraichnan turbulence (left versus right panels), where for a given magnetic field strength, the $L_{\rm c}$ that best fits the data is slightly smaller for  Kolmogorov than for Kraichnan turbulence (e.g. red and green lines for $B_{\rm rms}$=4\,$\mu$G).

\begin{figure}
\begin{center}
\includegraphics[width=0.48\textwidth]{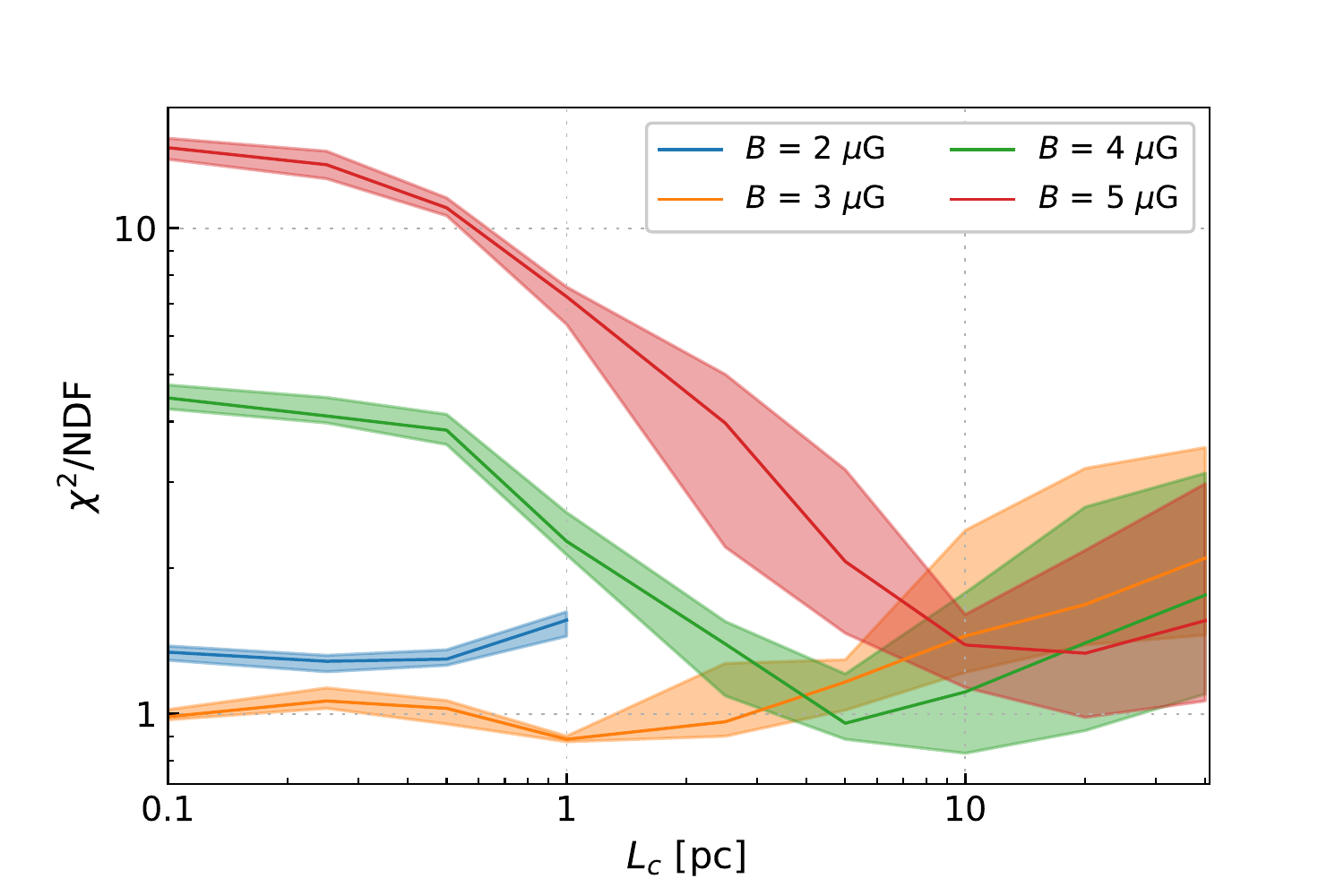}
\includegraphics[width=0.48\textwidth]{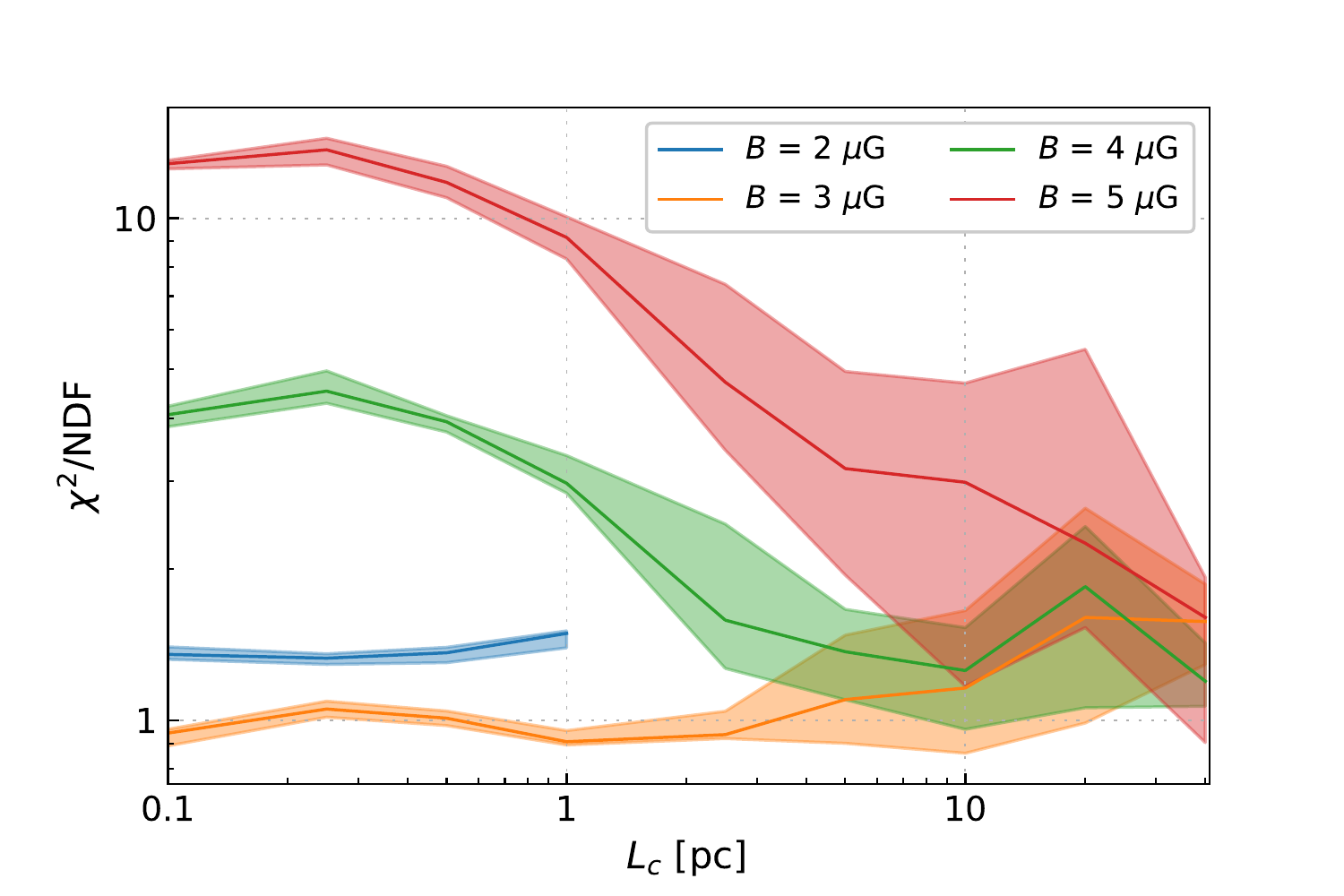}
\caption{$\chi^{2}$/ndf as a function of $L_{\rm c}$ for different magnetic field strengths. The top panel corresponds to Kolmogorov turbulence, and the bottom one to Kraichnan turbulence. Thick solid lines for the median $\chi^{2}$/ndf. Coloured bands: Between the 18th and 82th percentiles.}
\label{fig:chi2}
\end{center}
\end{figure}

To calculate the goodness of fit of these models to the data, we perform a $\chi^2$ fit of the model to the data for every realization performed. This gives us a distribution of $\chi^2$ values for every combination of turbulence parameters ---power spectrum, magnetic field strength and coherence length. To avoid averaging over outlier values that might result in a worsening of the fit parameters, we calculate the median of the $\chi^2$. We plot it in Figure~\ref{fig:chi2} (thick solid lines) and include it in Tables~\ref{tab:chi2:kol} and~\ref{tab:chi2:kra}. The confidence intervals of these distributions are shown as coloured bands between their 18th and 82th percentiles. For both Kolmogorov and Kraichnan turbulences, the overall best fit value is reached for a magnetic field of $B_{\rm rms}=3\,\mu$G and a coherence length of $L_{\rm c}\simeq 1\,$pc. For $B_{\rm rms}$=$2\,\mu$G the minimum of the $\chi^2$ is reached for $L_{\rm c}<$1\,pc. For Kraichnan turbulence, the best fit value for $B_{\rm rms}=4$ and $5\,\mu$G are reached for $L_{\rm c}$=10 and 40\,pc respectively. The best fit values for Kolmogorov turbulence for $B_{\rm rms}=4$ and $5\,\mu$G are reached for $L_{\rm c}=5$ and 10\,pc respectively. These minimum $\chi^2$ values $B_{\rm rms}=4$ and $5\,\mu$G are not far from that reached for $B_{\rm rms}=3\,\mu$G, specially for Kolmogorov turbulence. Nevertheless, the confidence bands are wider as a result of a larger scattering in the $\chi^2$ values. This is a consequence of the surface brightness having more irregular shapes due to the anisotropic diffusion in these regimes.

\begin{table}
\caption{$\chi^{2}$/ndf as a function of $L_{\rm c}$ and $B_{\rm rms}$, for Kolmogorov turbulence} 
\label{tab:chi2:kol}
\begin{center}
\begin{tabular}{ c|c|c|c|c| }
  Kolmogorov & 2\,$\mu$G & 3\,$\mu$G & 4\,$\mu$G & 5\,$\mu$G \\
 \hline
 0.1\,pc        &  20.1/15   &  14.8/15   &  67.1/15   &  220/15   \\
 0.25\,pc       &  19.2/15   &  15.9/15   &  61.6/15   &  203/15   \\
 0.5\,pc        &   19.5/15  &  15.4/15   &   57.6/15  &   165/15  \\
 1\,pc          &   23.4/15  &  13.3/15   &  34.0/15   &   108/15  \\
 2.5\,pc        &  N/A        &  14.4/15   &  20.9/15   &   59.6/15  \\
 5\,pc          &   N/A       &  17.5/15   &   14.4/15  &  30.9/15   \\
 10\,pc         &  N/A        &  21.7/15   &   16.7/15  &   20.8/15  \\
 20\,pc         &  N/A        &  25.2/15   &   21.0/15  &   20.0/15  \\
 40\,pc         &  N/A        &  31.4/15   &   26.4/15  &   23.4/15  \\
 \hline
\end{tabular}
\end{center}
\end{table}

\begin{table}
\caption{$\chi^{2}$/ndf as a function of $L_{\rm c}$ and $B_{\rm rms}$, for Kraichnan turbulence} \label{tab:chi2:kra}
\begin{center}
\begin{tabular}{ c|c|c|c|c| }
  Kraichnan & 2\,$\mu$G & 3\,$\mu$G & 4\,$\mu$G & 5\,$\mu$G \\
 \hline
 0.1\,pc        &  20.3/15   &  14.1/15   &  61.0/15   &  193/15   \\
 0.25\,pc       &  20.0/15   &  15.8/15   &  67.9/15   &  205/15   \\
 0.5\,pc        &   20.5/15  &  15.1/15   &   59.0/15  &   177/15  \\
 1\,pc          &   22.4/15  &  13.6/15   &  44.5/15   &   137/15  \\
 2.5\,pc        &  N/A        &  14.1/15   &  23.8/15   &   70.8/15  \\
 5\,pc          &   N/A       &  16.5/15   &   20.5/15  &  47.6/15   \\
 10\,pc         &  N/A        &  17.4/15   &   18.8/15  &   44.7/15  \\
 20\,pc         &  N/A        &  24.0/15   &   27.7/15  &   33.8/15  \\
 40\,pc         &  N/A        &  23.6/15   &   18.0/15  &   24.0/15  \\
 \hline
\end{tabular}
\end{center}
\end{table}

\begin{figure*}
\begin{center}
\includegraphics[width=0.48\textwidth]{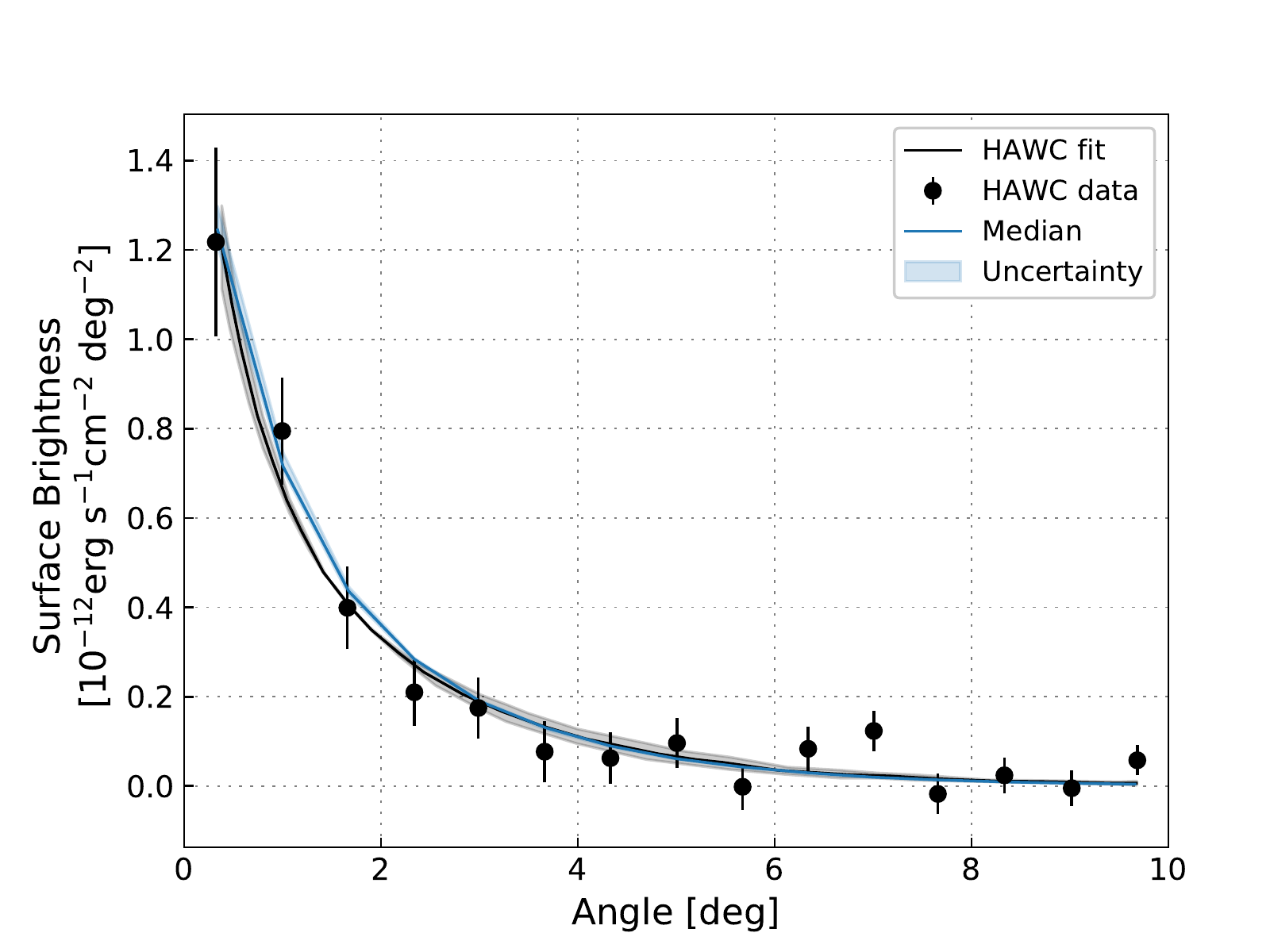}
\includegraphics[width=0.48\textwidth]{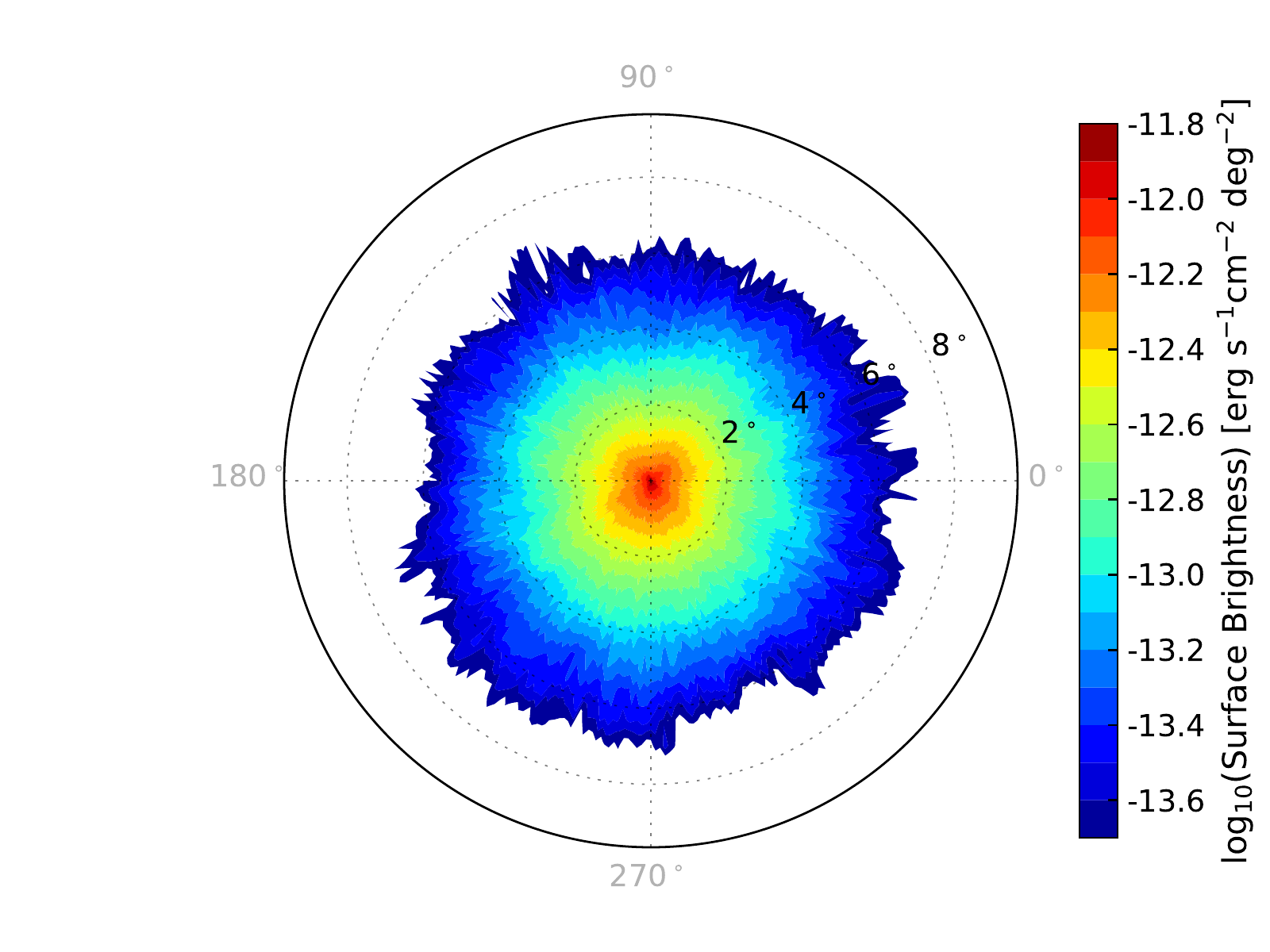}
\caption{Best fit results for Kolmogorov turbulence, with $B_{\rm rms}=3\,\mu$G and $L_{\rm c}=1$\,pc. Left panel: VHE $\gamma$-ray surface brightness as a function of the angular distance from the pulsar integrated over all the azimuthal angles. Right panel: VHE $\gamma$-ray surface brightness in polar coordinates. The radial distance indicates the angular distance from the pulsar and is marked with black ticks. The polar angle is marked with grey ticks.}
\label{fig:bestfit:kol}
\end{center}
\end{figure*}

The integrated surface brightness and that represented in polar coordinates for the best fit value of Kolmogorov turbulence ($B_{\rm rms}=3\,\mu$G and $L_{\rm c}=1$\,pc) are shown in Figure \ref{fig:bestfit:kol}. In the left panel we show in blue the best fit result as a median of all the electron realizations and the uncertainty as the 18th and 82th percentiles of the radial profile distributions. The black line shows the fit derived by the HAWC Collaboration in \cite{Geminga_hawc}. Apart from the qualitative agreement between both results, the best fit value of $\chi^2$/ndf = 13.3/15 shows that our model gives a very good fit to the data provided by HAWC. To evaluate the asymmetry of the $\gamma$-ray profile, we also show the two dimensional profile of the gamma-ray spectrum generated by the simulated electrons on the right panel of Figure~\ref{fig:bestfit:kol}. This spatial distribution is similar to that shown by HAWC on the Test Statistic skymap they show in Figure 1 of \cite{Geminga_hawc}.

\section{Discussion}
\label{Discussion}

Propagating individual electrons in 3D realizations of magnetic turbulence, we have demonstrated that the extended gamma-ray emission detected by HAWC around Geminga is compatible with the emission from $\sim 100$\,TeV electrons in isotropic, homogeneous Kolmogorov or Kraichnan turbulence with realistic physical parameters. The best fit values for Kolmogorov turbulence are found to be $L_{\rm c} \approx 1$\,pc and $B_{\rm rms} \approx 3\,\mu$G, although reasonably larger values of $L_{\rm c}$ and $B_{\rm rms}$ are possible. Kraichnan turbulence favours a slightly larger $L_{\rm c} \approx 2$\,pc for $B_{\rm rms} = 3\,\mu$G, but does not give subtantially different results. As should be expected, larger coherence lengths are preferred for stronger magnetic fields, cf. Fig.~\ref{fig:chi2}. However, too weak ($B_{\rm rms} \lesssim 2\,\mu$G) or too strong ($B_{\rm rms} \gtrsim 5\,\mu$G) magnetic fields all start to give a bad fit to the data, at least within the two isotropic turbulence models we considered here. In particular, large values of $B_{\rm rms}$ require too large coherence lengths, leading to unrealistically large asymmetries in the gamma-ray emission around Geminga. Such large asymmetries are not supported by the current HAWC data. The limit on the maximum coherence length would change if the VHE gamma-ray emission from Geminga turns out to be asymmetric, but coherence lengths larger than the size of the source can be completely ruled out because they would produce morphologies as the one shown in the lower right panel of Figure~\ref{fig:2Dprofiles}.

It is interesting to note that our favoured values for $L_{\rm c}$ and $B_{\rm rms}$ are fully consistent with the ranges of values measured in the disc of our Galaxy with radio observations, even though they lie somewhat closer to the lower end of these ranges. In the Milky Way, $L_{\rm c}$ is typically $\sim 1 - {\rm a~few~} 10$s~pc~\citep{Haverkorn:2008tb,Iacobelli2013}, and $B_{\rm rms}$ is measured between a few $\mu$G and $\sim 10\,\mu$G. See e.g. \cite{JF1,JF2} for a recent Galactic magnetic field model.

The fact that \lq\lq typical\rq\rq\/ turbulence parameters fit the data is also interesting in light of the large discrepancy between HAWC measurement of the CR diffusion coefficient, and the value that is usually inferred from the boron-to-carbon ratio \citep{BC_AMS,GALPROP} and commonly used within the CR community. First, this gives credence to the possibility that the turbulence probed by these $\gamma$-ray emitting electrons around Geminga is not very significantly different from the interstellar turbulence elsewhere in the Galactic disc. Second, this provides another independent argument in favour of a CR diffusion coefficient in the disc smaller than usually thought. A two-orders-of-magnitude smaller diffusion coefficient is preferred, cf. \cite{Geminga_hawc} and Section~\ref{Results} of the present study. Let us however point out that regions with larger coherence lengths, and thence faster CR diffusion at fixed $B_{\rm rms}$, probably exist in the disc. For example, \cite{Frisch:2012zj,Frisch:2015hfa} claim that the local interstellar magnetic field around the Earth is coherent over a few tens of parsecs, which would favour $L_{\rm c} \geq 10$\,pc in our local environment. Nonetheless, even for such larger values of $L_{\rm c}$, the CR diffusion coefficient in pure Kolmogorov or Kraichnan turbulence would still be significantly smaller than that inferred from the B/C ratio by a factor of several tens, see Figure~\ref{fig:diff_coef} in Section~\ref{D_at_100TeV}. In that sense, even assuming that $L_{\rm c}$ around Geminga is unusually small would not solve the aforementioned discrepancy. Several studies presenting numerical calculations of the CR diffusion coefficient in isotropic turbulence have already found it to be smaller than that inferred from the B/C ratio by a factor of at least a few tens~\citep{Casse2002,DeMarco2007,2012PhRvL.108z1101G,2013PhRvD..88b3010G}. Our calculations in Section~\ref{D_at_100TeV} are consistent with the findings from these previous papers. These low-diffusion regions have previously been observed, as for example in the Cygnus superbubble~\citep{Ackermann11b}, or around the supernova remnant W28~\citep{Gabici2010}. The two latter measurements may however probe regions that are not representative of the interstellar medium far from any source: CRs escaping from their sources could drive instabilities that would locally lower the diffusion coefficient value around them, see e.g.~\cite{Malkov2013,DAngelo2016,Nava2016}. \cite{Geminga_hawc} argued that such a limitation should not hold for the case of the diffusion coefficient measured by HAWC around Geminga, because of the low energy density of the electrons responsible for the $\gamma$-ray emission, compared to the energy density of the interstellar medium.


Even though it is currently impossible to conclude that CRs diffuse so slowly in the entire (thin) Galactic disc, the possibility of a smaller diffusion coefficient deserves more attention. Such a scenario is {\it a priori} not impossible, in that standard estimates of the diffusion coefficient from the boron-to-carbon ratio are made within the confines of idealized models.

The fact that the simplest models of isotropic turbulence are found to fit well HAWC data does not discard the possibility that interstellar turbulence is anisotropic. Indeed, the gyroradius $r_{\rm g}$ of $\sim 100$\,TeV electrons is only an order of magnitude smaller than the preferred coherence length $L_{\rm c} \sim 1$\,pc. This implies that the electrons responsible for the emission measured by HAWC only probe the part of the turbulence spectrum close to the outer scale. For instance, Goldreich-Sridhar turbulence~\citep{SridharGoldreich1994,Goldreich:1994zz} is strongly anisotropic only at large wave vectors, and is more isotropic close to the outer scale.

The reason why HAWC data currently does not allow for a clear separation between a Kolmogorov and a Kraichnan spectrum is similar. As can be seen by comparing the left-hand sides of both panels in Fig.~\ref{fig:diff_coef}, the CR diffusion coefficient normalization is almost identical when $r_{\rm g} \rightarrow L_{\rm c}$. The difference between the two types of turbulence only becomes noticeable when $r_{\rm g} \ll L_{\rm c}$: See the right-hand sides of both panels in Fig.~\ref{fig:diff_coef} for results at larger $L_{\rm c}$, at $r_{\rm g}$ fixed. In practice, $L_{\rm c}$ is fixed and one would need to study lower-energy electrons, and thence detect lower-energy gamma rays in that region, so as to distinguish between the two power-spectra. Alternatively, dividing HAWC emission into at least two energy bins may give some insights.

Finally, the present work shows how extended gamma-ray emission can be used as a novel way to probe interstellar magnetic fields around CR sources. The advantage of this type of studies compared with those of hadronic gamma rays from molecular clouds near CR sources is the possibility to gain two-dimensional information on the distribution of emitting particles around the source. If a molecular cloud sits at a distance $\lesssim L_{\rm c}$ from a CR source, then the deduced CR diffusion coefficient can differ by several orders of magnitude from its typical value on scales $\gtrsim L_{\rm c}$~\citep{2013PhRvD..88b3010G}: It depends on whether the cloud is directly connected to the source by magnetic field lines or not.

\section{Conclusions and perspectives}
\label{Conclusions}

In this paper, we placed constraints on the properties of the turbulent magnetic fields in the Geminga region, using HAWC VHE $\gamma$-ray data. We tested two types of turbulence (Kolmogorov and Kraichnan), and probed a wide range of parameter values. For both turbulence models, we find that the surface brightness measured by HAWC is best reproduced with a magnetic field strength at the level of $\approx 3\,\mu$G and a coherence length at the level of $L_{\rm c} \approx 1$\,pc. Weaker or significantly stronger magnetic fields do not give a good fit to the data. Substantially larger coherence lengths ($L_{\rm c} \gtrsim 10$\,pc) are disfavoured because they would result in a large asymmetry of the VHE $\gamma$-ray emission, see e.g. the extreme case of $L_{\rm c} = 40$\,pc in the lower right panel of Figure~\ref{fig:2Dprofiles}. No significant asymmetry was reported by HAWC in this region. Our numerical calculations of the CR diffusion coefficient in pure turbulence (no regular field) yield values at 100\,TeV that are compatible with that measured by HAWC.

At the present time, it is difficult to constrain the power-spectrum of the turbulence. Kolmogorov and Kraichnan turbulence both give similar results at these $\gamma$-ray energies. In order to establish the energy-dependence of the CR diffusion coefficient at these energies, and thence constrain the power-spectrum, it is important for future $\gamma$-ray measurements to be able to provide an energy-dependent morphology of the sources. This can be reached with an improved reconstruction and energy resolution by the HAWC Observatory, currently taking data on the source, or by scanning the region using imaging atmospheric Cherenkov telescopes with an improved sensitivity and larger FoV cameras as the ones that will be installed in the future CTA~\citep{CTA_Concept}.

\section*{Acknowledgements}
The authors would like to thank Jim Hinton for useful discussions and help during the production of the paper. The authors would also like to thank the HAWC collaboration for useful discussions. The research of GG was supported by a Grant from the GIF, the German-Israeli Foundation for Scientific Research and Development.

\bibliographystyle{./mnras} 

 \bibliography{./references}

\bsp	
\label{lastpage}
\end{document}